\documentclass[aps,prl,twocolumn,superscriptaddress,amsmath,amssymb]{revtex4}
\usepackage{graphicx}
\usepackage{epstopdf}
\usepackage{longtable}
\usepackage{hyperref}
\usepackage{amssymb}
\usepackage{dcolumn}
\usepackage{bm}

\begin{document}

\title{Anisotropic Lattice Compression and Pressure-Induced Electronic Phase Transitions in Sr$_2$IrO$_4$}

\author{K. Samanta}
\affiliation{``Gleb Wataghin'' Institute of Physics, University of Campinas - UNICAMP, Campinas, S\~ao Paulo 13083-859, Brazil}

\author{R. Tartaglia}
\affiliation{``Gleb Wataghin'' Institute of Physics, University of Campinas - UNICAMP, Campinas, S\~ao Paulo 13083-859, Brazil}

\author{U. F. Kaneko}
\affiliation{Brazilian Synchrotron Light Laboratory (LNLS), Brazilian Center for Research in Energy and Materials (CNPEM), Campinas, S\~ao Paulo 13083-970, Brazil}

\author{N. M. Souza-Neto}
\affiliation{Brazilian Synchrotron Light Laboratory (LNLS), Brazilian Center for Research in Energy and Materials (CNPEM), Campinas, S\~ao Paulo 13083-970, Brazil}

\author{E. Granado}
\affiliation{``Gleb Wataghin'' Institute of Physics, University of Campinas - UNICAMP, Campinas, S\~ao Paulo 13083-859, Brazil}

\begin{abstract}

The crystal lattice of Sr$_2$IrO$_4$ is investigated with synchrotron X-ray powder diffraction under hydrostatic pressures up to $P=43$ GPa and temperatures down to $20$ K. The tetragonal unit cell is maintained over the whole investigated pressure range, within our resolution and sensitivity. The $c$-axis compressibility $\kappa_c(P,T) \equiv -({1} / {c}) ({d c} / {d P})$ presents an anomaly with pressure at $P_1=17$ GPa at fixed $T=20$ K that is not observed at $T=300$ K, whereas $\kappa_a(P,T)$ is nearly temperature-independent and shows a linear behavior with $P$. The anomaly in $\kappa_c(P,T)$ is associated with the onset of long-range magnetic order, as evidenced by an analysis of the temperature-dependence of the lattice parameters at fixed $P=13.7 \pm 0.5$ GPa. At fixed $T=20$ K, the tetragonal elongation $c/a(P,T)$ shows a gradual increment with pressure  and a depletion above $P_2=30$ GPa that indicates an orbital transition and possibly marks the collapse of the $J_{eff}=1/2$ spin-orbit-entangled state. Our results support pressure-induced phase transitions or crossovers between electronic ground states that are sensed, and therefore can be probed, by the crystal lattice at low temperatures in this prototype spin-orbit Mott insulator.

\end{abstract}

\maketitle

\section{Introduction}

\begin{figure}[t]
	\includegraphics[width=0.48\textwidth]{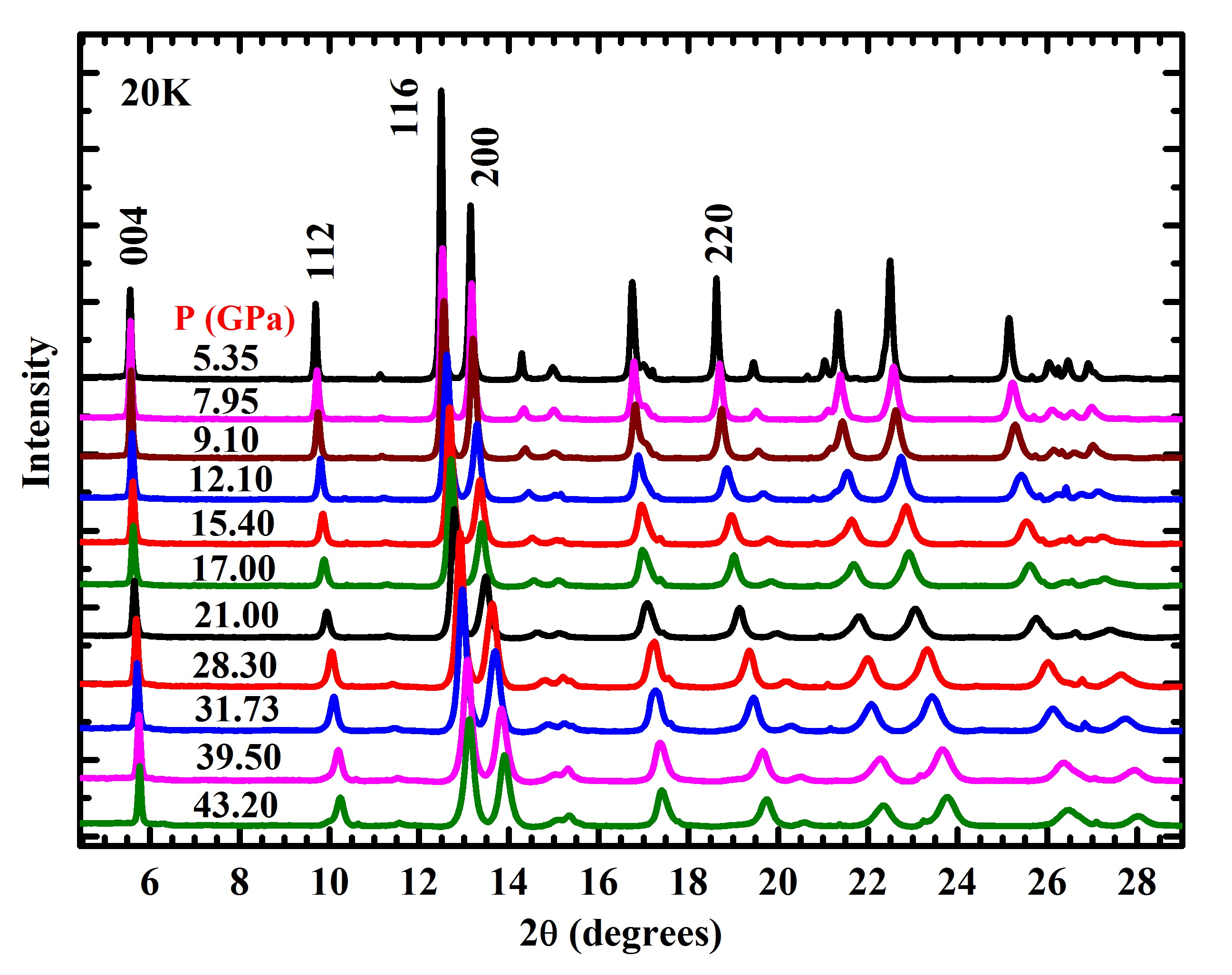}
	\caption{\label{raw} Raw X-ray powder diffraction profiles for several pressures at $T=20$ K ($\lambda=0.61986$ \AA). The Miller indices of selected reflections are indicated.}
\end{figure}

\begin{figure*}[t]
	\includegraphics[width=0.8\textwidth]{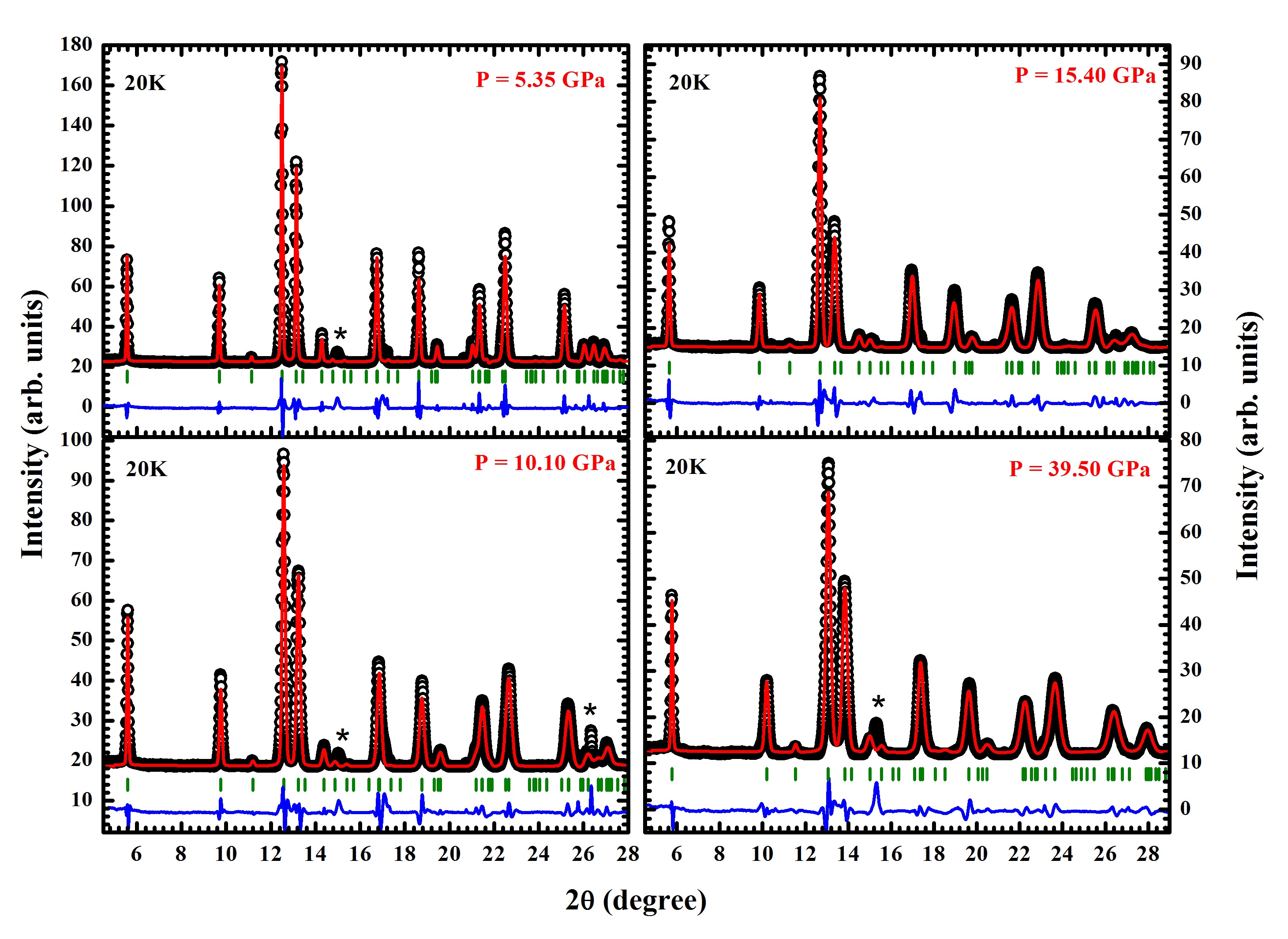}
	\caption{\label{rietveld} Comparison between observed (open symbols) and calculated (red lines) X-ray powder diffraction profiles for selected pressures and $T=20$ K. The differences are also shown (blue lines). The expected Bragg peak positions for the tetragonal  $I4_1/acd$ space group are shown in short vertical bars. The spurious peaks identified by `*' are due to the rhenium gasket.}
\end{figure*}

\begin{figure}[t]
	\includegraphics[width=0.45\textwidth]{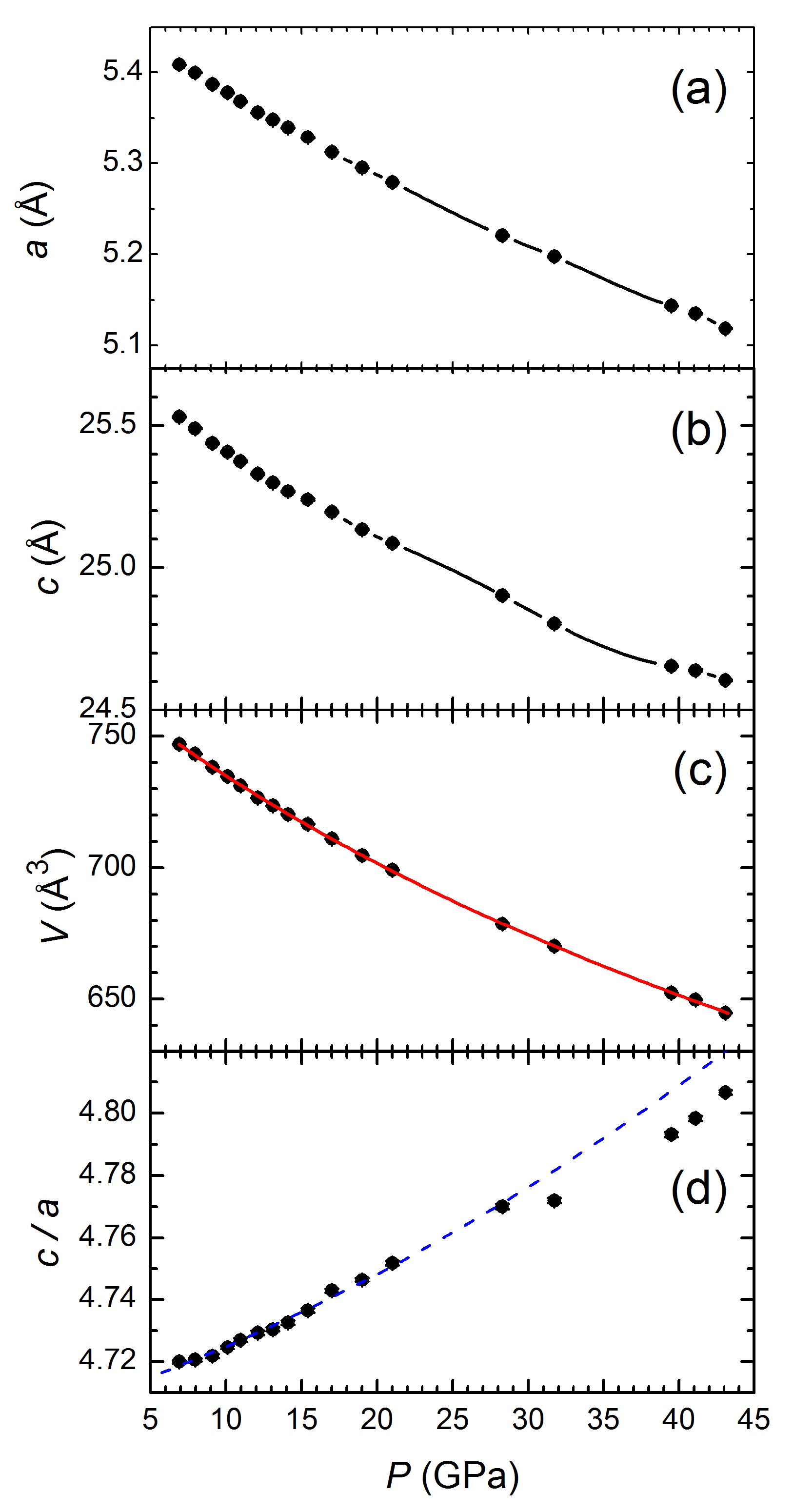}
	\caption{\label{lattpar} Pressure-dependence of the tetragonal lattice parameters $a$ (a), $c$ (b), unit cell volume $V$ (c), and $c/a$ ratio (d) at $T=20$ K. The red solid line in (c) is a fitting according to the Murnaghan equation of state (see text). The blue dashed line in (d) is a second-order polynomial curve that captures the behavior below 30 GPa, and is shown as a guide to the eyes.}
\end{figure}

\begin{figure}[t]
	\includegraphics[width=0.45\textwidth]{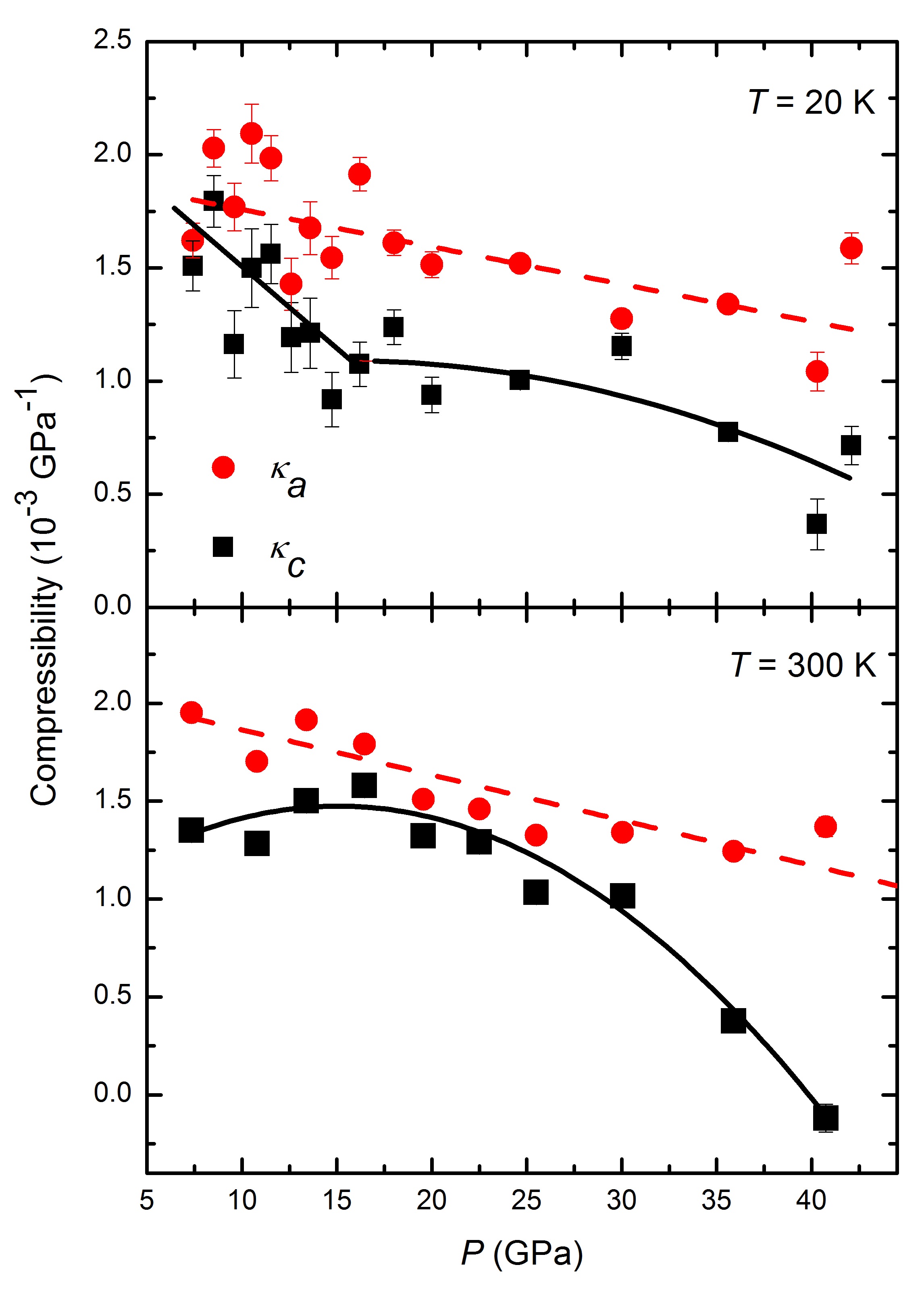}
	\caption{\label{compress} Pressure-dependence of the $a$-axis compressibility, $\kappa_a \equiv -({1} / {a}) ({d a} / {d P})$ and the $c$-axis compressibility, $\kappa_c \equiv -({1} / {c}) ({d c} / {d P})$ at $T = 20$ K (a) and room temperature (b). The data in (b) were extracted from the results of Ref. \onlinecite{Samanta}. The solid and dashed lines are guides to the eyes.}
\end{figure}

\begin{figure}[t]
\includegraphics[width=0.45\textwidth]{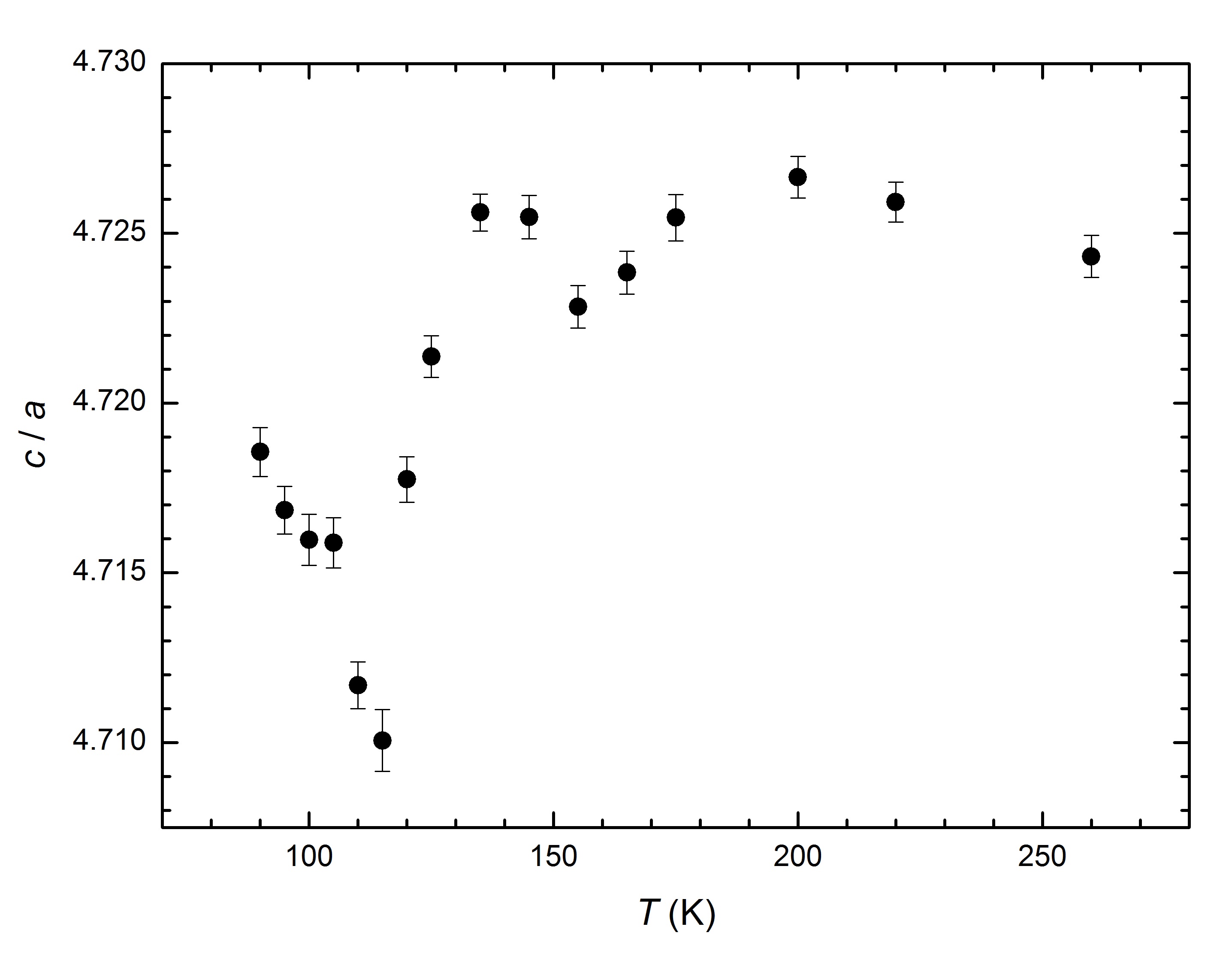}
	\caption{\label{elongation} Temperature-dependence of the tetragonal elongation ($c/a$ ratio) on warming under constant $P=13.7 \pm 0.5$ GPa.}
\end{figure}

The effect of strong spin-orbit coupling in the physical properties of solids is an active area of research in contemporary condensed matter physics \onlinecite{Kane,Bernevig,Konig,Pesin}. The $5d$-transition metal oxides, in particular the Ruddlesden-Popper series of strontium iridates Sr$_{n+1}$Ir$_n$O$_{3n+1}$, are of interest due to the rare combination of strong spin-orbit interaction and electronic correlations that leads to a spin-orbit-entangled Mott insulating state ($J_{eff} = 1/2$) \onlinecite{Kim1,Kim2,Rau,Caorev,Arita,Li}. According to powder diffraction experiments, the layered crystal structure of Sr$_2$IrO$_4$ ($n=1$) can be well described under a tetragonal space group $I4_1/acd$, featuring a rotation of the IrO$_6$ octahedra by $\sim 11 ^{\circ}$ along the $c$-axis \onlinecite{Huang,Crawford}. A significant Dzyaloshinskii-Moriya-type (DM) antisymmetric exchange interaction leads to a canted magnetic structure with a weak ferromagnetic moment of $0.06-0.14$ $\mu_B$/Ir below $T_N = 240$ K at ambient pressure \onlinecite{Crawford,Chikara,Cao1}. The in-plane magnetic canting angle is nearly locked to the octahedral rotation  \onlinecite{Kim1,Kim2,Ye,Jackeli}. On the other hand, the $J_{eff} = 1/2$ state state may be unstable against a slight distortion of the IrO$_6$ octahedra due to possible competing orbital configurations \onlinecite{Jackeli,Lado}. Application of external pressure is expected to induce changes in the interatomic distances and angles, which could tune such competing electronic ground states. It is also generally expected that external pressure should favor a metallic against a Mott insulating state, since the compressed lattice should increase the hybridization of the atomic wavefunctions thereby increasing the bandwidths. However, a metallic state is not achieved up to at least 55 GPa in Sr$_2$IrO$_4$, and the resistivity displays a U-shaped curve with pressure \onlinecite{review,Haskel,Zocco}.
Our recent combined synchrotron X-ray powder diffraction (XPD) and phonon Raman scattering study under pressure at room temperature revealed a first-order structural phase transition at $\sim 50$ GPa, and anomalous phonon and lattice behavior below 40 GPa that are indicative of electronic instabilities \onlinecite{Samanta}. On the other hand, it is desirable to extend the detailed crystallographic investigation of this system under pressure to the low-temperature region of the phase diagram, so the relevance of the electronic/magnetic ground states to the crystal lattice can be probed. Here, we report a detailed XPD study in Sr$_2$IrO$_4$ under hydrostatic pressures up to $P=43$ GPa and down to $T=20$ K. The long-range magnetic order is shown to produce an increment of the c-axis compressibility $\kappa_c(P,T) \equiv -({1} / {c}) ({d c} / {d P})$, as revealed both by pressure-dependence (at low temperature)  and temperature-dependence (at $P=13.7 \pm 0.5$ GPa) of the lattice parameters. Also, the $c/a$ ratio shows a depletion consistent with an orbital transition above $P_2=30$ GPa. Our results demonstrate that pressure-induced changes in the electronic and/or magnetic ground state of Sr$_2$IrO$_4$ have a measurable impact in the crystal lattice at low temperatures, and, conversely, such lattice anomalies can be used to investigate the nature of the electronic/magnetic transitions of this and related materials.



\section{Experimental Details}

The Sr$_2$IrO$_4$ powder sample was synthesized by standard high-temperature solid-state reaction mechanism as described in Ref. \onlinecite{Samanta}. Pressure-dependent XPD measurements at low temperature were carried out at the X-ray Diffraction and Spectroscopy (XDS) beamline of the Brazilian Synchrotron Light Laboratory (LNLS) \onlinecite{XDS} with $\lambda = 0.61986$ \AA\ calibrated with a LaB$_6$ standard sample. The focal spot size is $90 \times 40$ $\mu m^2$, obtained with a Rh cylindrical mirror, a $L$N$_2$-cooled double flat Si(111) crystal monochromator, a Rh toroidal focusing mirror, and a Kirkpatrick-Baez mirror. The diffracted Debye rings were detected in transmission geometry by a 2D-detector and the ring intensities were integrated to yield conventional $I$ versus $2 \theta$ diffractograms. A CuBe diamond anvil cell (DAC) was placed inside a continuous-flow liquid He cryostat. Boehler-Almax-type ultralow fluorescence diamonds with a cullet diameter of 350 $\mu$m were used to generate pressure. The pressure transmitting medium was helium, and gaskets were made of rhenium. The pressure values were obtained using the well-known ruby $R_1$ fluorescence lineshift method. Pressure inside the DAC was controlled using a gas membrane. The pressure-dependent measurements were taken under increasing pressurization after the target temperature was stabilized. Temperature-dependent X-ray diffraction data at fixed pressure ($13.7 \pm 0.5$ GPa) were collected under warming after cooling the cell with residual pressure ($<5$ GPa) and applying the desired pressure at the base temperature. The target pressure was checked at each temperature. For each measurement run (either pressure- or temperature-dependent), a new virgin sample was loaded in the pressure cell.

\section{Results and Analysis}

Raw XPD profiles of Sr$_2$IrO$_4$ up to $P=43$ GPa and at $T=20$ K are shown in Fig. \ref{raw}. Bragg peaks consistent with a tetragonal space group $I4_1/acd$ are observed in this pressure range, within our resolution and sensitivity. Rietveld refinements were performed in order to extract the lattice parameters of this phase under pressure. Since the accuracy of the Bragg intensities was limited by poor grain statistics (see ref. \onlinecite{Samanta}) and possible preferred orientation effects, the atomic positions and Debye Waller factors were kept fixed in the refinements at previously reported values at zero pressure (Ref. \onlinecite{Crawford}). Figure \ref{rietveld} shows the full  observed and calculated diffraction profiles at selected pressures, revealing sufficiently good fits that led to reliable refined lattice parameters.

The pressure-dependence of the $a$ and $c$ latice parameters at 20 K are shown in Figs. \ref{lattpar}(a) and \ref{lattpar}(b), respectively. The corresponding unit cell volume $V$ is given in Fig. \ref{lattpar}(c) (symbols). A  Murnaghan equation of state \onlinecite{Murnaghan} $P=(B_0/B_0')[(V_0/V(P))^{B_0'}-1])$ yielded a satisfatory fit (solid line in Fig. \ref{lattpar}(c)), with the initial bulk modulus $B_0=162(3)$ GPa, initial bulk modulus derivative $B_0'=3.6(2)$ and initial unit cell volume $V_0=777.0(8)$ \AA$^3$. The tetragonal elongation ($c/a$ ratio) is given in Fig. \ref{lattpar}(d), increasing under pressurization. A clear anomaly in the $c/a$ curve is seen at $P_2=30$ GPa.

Figures \ref{compress}(a) and \ref{compress}(b) show the $a$- and $c$-axes compressibilities $\kappa_a(P)$ and $\kappa_c(P)$ obtained from the refined lattice parameters at (a) 20 K and (b) room temperature, respectively, where the latter was extracted from the data shown Ref. \onlinecite{Samanta}. It can be seen that the $\kappa_a(P)$ curve is similar at both temperatures, showing a smooth reduction under increasing pressures, whereas $\kappa_c(P)$ is sensitive to temperature. In fact, at 20 K, $\kappa_c(P)$ decreases from $1.7(1) \times 10^{-3}$ GPa$^{-1}$ at 7 GPa to $1.0(1) \times 10^{-3}$ GPa$^{-1}$ at $P_1 \equiv 17$ GPa, remaining nearly constant at this value up to $P_2 \equiv 30$ GPa and further decreasing at higher pressures. At room temperature, $\kappa_c(P)$ increases slightly from $1.2(1) \times 10^{-3}$ GPa$^{-1}$ at 7 GPa to  $1.6(1) \times 10^{-3}$ GPa$^{-1}$ at $P_1$, then decreasing to $1.0(1) \times 10^{-3}$ GPa$^{-1}$ at $P_2$ and steeply approaching to zero at higher pressures.

In order to investigate whether the different $\kappa_c(P)$ curves at $T=20$ K and $T=300$ K may be related to a phase transition, more detailed temperature-dependent measurements are necessary. Since $\Delta c(P) = -\int_0 ^P \kappa_c(P') dP'$, the increased $c$-axis compressibility at low temperatures and low pressures would be consistent with a larger integrated $c$-axis compression $-\Delta c(P)$ at 20 K with respect to room $T$ in the low-pressure limit. In order to further investigate this effect, we obtained the temperature-dependence of the lattice parameters on warming while keeping a nearly constant applied pressure $13.7 \pm 0.5$ GPa. Whereas the slight pressure variations for each change of temperature are sufficient to induce significant non-statistical fluctuations for both $a$ and $c$ (not shown), the $c/a$ ratio is much less sensitive to such pressure variations, yielding a physically meaningful temperature-dependence (see Fig. \ref{elongation}). This ratio decreases steady on warming up to $\sim 115$ K, shows a significant increment above this temperature, and finally stabilizes at a constant value $c/a=4.725$ above 130 K. Considering that $\kappa_a(P)$ is temperature-independent within our resolution, the increment of $c/a (P=13.7$ GPa) above 115 K shown in Fig. \ref{elongation} is consistent with the larger $\kappa_c(P)$ at low temperatures and low pressures with respect to room temperature and low pressures (see Fig. \ref{compress}). 

\section{Discussion}

The $\kappa_c(P)$ curves displayed in Figs. \ref{compress}(a) and \ref{compress}(b) indicate a change of behavior at $P_1=17$ GPa. Also, a clear anomaly in the $c/a$ ratio is observed at $P_2=30$ GPa (see Fig. \ref{lattpar}(d)). Previous Raman scattering data at room temperature show significant phonon anomalies at both $P_1$ and $P_2$ (Ref. \onlinecite{Samanta}). Also, the magnetic XMCD signal and antiferromagnetic diffraction peaks at low temperatures disappear at $P_1$, whereas the XAS $L_3/L_2$ branching ratio appears to show a significant reduction above $P_2$ (Refs. \onlinecite{Haskel} and \onlinecite{Haskel2}). Taking all these independent data into consideration, there is compelling evidence of phase transitions or crossovers at these critical pressures, which are manifested at room temperature as well as at low temperatures. In short, the observed pressure-induced anomalies in the crystal lattice are likely related to modifications in the electronic structure of this material.

It can be seen in Figs. \ref{lattpar} and \ref{compress} that the lattice compression is anisotropic, consistent with previous data \onlinecite{Haskel,Samanta}. In fact, since $\kappa_a > \kappa_c$ for all pressures, the tetragonal elongation $c/a$ increases with $P$ [see Fig. \ref{compress}(d)]. This is a reasonable trend, since the compression in the $ab$-plane may occur both by a reduction of bond distances and by an increment of the tilt angle of the IrO$_6$ octahedra along the $c$-axis, whereas only the first of these mechanisms is operative for the $c$-axis compression. Up to $P_2$, the increment of $c/a$ ratio is well captured by a second-order polynomial [dashed line in Fig. \ref{compress}(d)]. However, above $P_2$ the $c/a$ ratio clearly falls below the extrapolated polynomical behavior. For pressures significantly above $P_2$, this ratio appears to restablish the same increment rate as the extrapolated polynomial, however with a constant negative offset. This behavior is suggestive of an orbital transition at $P_2$ with an increment of Ir $5d$ electronic density in the $ab$ plane above this pressure. Since the $5d$ electronic density in the ideal spin-orbit-entangled $J_{eff}=1/2$ state shows cubic symmetry with equal $xy$, $xz$ and $yz$ orbital occupations, the reported anomaly in the $c/a$ ratio at $P_2$ is indicative of a breakdown of such state. A possible scenario that is consistent with our data is a transition from a $J_{eff}=1/2$ state at low pressures to a state with fully occupied $xy$ orbitals and a hole in a combination of $xz$ and $yz$ orbitals above $P_2$. In fact, ab-initio electronic calculations indicate that such state may be energetically viable with repect to the $J_{eff}=1/2$ configuration \onlinecite{Lado}. 

The structural anomaly observed at $P_1$ appears to be more subtle with respect to that at $P_2$. In fact, no clear jump is observed in $c/a$, suggesting that Ir$^{4+}$ orbital configuration is not substantially altered at $P_1$. This is consistent with the nearly constant $L_3/L_2$ XAS branching ratio through $P_1$ \onlinecite{Haskel}. On the other hand, $\kappa_c$ shows a clear change of behavior at $P_1$ and low temperatures (see Fig. \ref{compress}(a)). Our temperature-dependent lattice compression data at $P=13.7$ GPa indicates that the increased compressibility at low pressures is related to a phase transition that occurs at $T^{*} \sim 115-130$ K (see Fig. \ref{elongation}), which coincides with $T_N$ for this pressure (see Ref. \onlinecite{Haskel2}). It is therefore clear that the increment of $\kappa_c$ at low pressures and low temperatures is associated with the presence of long-range magnetic order under these conditions. It is interesting to mention that the reduction of $c/a$ below $T^{*}$ at $P=13.7$ GPa contrasts with the behavior at ambient pressure, where $c/a$ increases below $T_c$ (Ref. \onlinecite{Bhatti}). Therefore, the anomalies in $c/a$ shown in Fig. \ref{elongation} most likely reflect the increment in $\kappa_c$ in the magnetically ordered phase rather than a magnetoelastic coupling at zero pressure.

It is not very surprising that the lattice compressibility is sensitive to magnetic order. In fact, the magnetic energy may modify the lattice spring constants directly through the spin-phonon coupling \cite{Baltensperger,Granado,Samanta2}, whereas less direct mechanisms involving the coupling of magnetic order to the electronic structure through the spin-orbit coupling might also affect the lattice stiffness. However, the sensitivity of $\kappa_c$ but not $\kappa_a$ to long-range magnetic order is interesting, considering that the exchange coupling in the $ab$ plane is much stronger than along $c$. This result is likely a consequence of the much larger magnetic correlation length within the $ab$ plane than along $c$ in the paramagnetic phase of this material \onlinecite{Fujiyama}, which may effectively wash out the sensitivity of $\kappa_a$ to the long-range magnetic ordering transition.


Finally, we should mention that anisotropic lineshape broadenings of Bragg peaks were previously reported at room temperature and for $P>P_1$ \onlinecite{Samanta}, which was interpreted as a possible sign of a symmetry-breaking instability. This line of reasoning led to the expectation that the crystal structure should suffer a long-range phase transition at $P_1$ for sufficiently low temperatures, which is however not confirmed by our present data at least at $T=20$ K. An alternative explanation for the anisotropic lineshape broadenings reported in ref. \onlinecite{Samanta} is based on the anisotropic compressibility of the crystal structure (see Fig. \ref{compress}). Spatial fluctuations of the stress may occur and become more pronounced as the applied pressure increases. This may result in anisotropic strain fluctuations due to the different values of $\kappa_a$ and $\kappa_c$, therefore leading to the anisotropic lineshape broadenings of the Bragg peaks. On the other hand, it is interesting to note that for Sr$_3$Ir$_2$O$_7$, which is another member of this Ruddlesden-Popper series with $n=2$, a tetragonal-monoclinic transition was observed at pressures close to 15 GPa \onlinecite{Zhang}, which is comparable to the critical pressure $P_1$ for Sr$_2$IrO$_4$. Further investigations are necessary to explore the similarities and differences between these related materials.


\section{Conclusions}

In summary, the lattice parameters of pressurized Sr$_2$IrO$_4$ shows a tetragonal elongation that is increased significantly by hydrostatic pressures of the order of tens of GPa, with anomalies in $\kappa_c$ at $P_1=17$ GPa at low temperatures and $c/a$ ratio at $P_2=30$ GPa. The anomaly at $P_2$ is interpreted in terms of a transition between competing orbital configurations of the Ir $5d$ $t_{2g}$ hole, whereas the anomaly of $\kappa_c$ at $P_1$ is associated with the onset of long-range magnetic order.

\section{acknowledgements}

We thank D. Haskel and G. Fabbris for illuminating discussions and for sharing unpublished data, and M. Eleot\'erio, J. Fonseca J\'unior and R.D. Reis for technical assistance. LNLS is acknowledged for concession of beamtime. This work was supported by Fapesp Grants 2016/00756-6, 2017/10581-1, and 2018/20142-8, and CNPq Grants 308607/2018-0 and 409504/2018-1, Brazil.

\end{document}